\documentstyle[aps,amssymb]{revtex}

\begin{document}
\title{Magnetic and thermodynamic properties of Sr$_{2}$LaFe$_{3}$O$_{9}$}
\author{H.Kachkachi$^{a}$, P.Kornilovitch$^{b}$, and Y.Meurdesoif$^{a}$}
\address{$^{a}$CPTMB, CNRS-URA 1537, University Bordeaux I, Rue du Solarium, 33174\\
Gradignan Cedex, France\\
$^{b}$King's College, Strand, London WC2R 2LS, UK.\\
\vspace{1cm}}
\author{J.-C. Grenier, and F. Zhou}
\address{ICMCB, CNRS, Avenue du Docteur A. Schweitzer, 33608 Pessac Cedex\\
France}
\maketitle

\begin{abstract}
We use a Dirac-Heisenberg Hamiltonian with biquadratic exchange interactions
to describe the first-order magnetic transition occurring in the perovskite
Sr$_{2}$LaFe$_{3}$O$_{9}$. Up on fitting the experimental curve for the
magnetic susceptibility below and above the N\'{e}el temperature, we give an
estimate of the exchange integrals for the antiferromagntic and
ferromagnetic interactions in this compound. Within linear spin-wave theory
we find that the magnon spectrum comprises a gapless antiferromagnetic mode
together with two gapped ferromagnetic ones.


{\bf keywords} : antiferromagnetic-ferromagnetic ordering, biquadratic
interactions, first-order magnetic transition.

\vspace{2cm}

Corresponding author: kachkach@bortibm4.in2p3.fr
\end{abstract}

\section{Introduction}

Since the discovery of high-T$_{c}$ copper-oxide superconductors\cite{BM},
investigations of superconductivity and magnetism of d-electron based oxides
with perovskite structure have gained new interest. Iron based oxides are
not superconducting, but exhibit interesting electronic properties such as
disproportionation as in CaFeO$_{3}$ or metalic properties as in SrFeO$_{3}$%
. In addition, these ferrites often shown very peculiar magnetic properties.

Almost stoichiometric Sr$_{2}$LaFe$_{3}$O$_{9-\delta }$, ($\delta \simeq 0$)
can be prepared under air by long annealing at $200{{}^\circ}C$ or using
electrochemical oxidation in alkaline solution\cite{battle}, \cite{JCGFZ}, 
\cite{WLM}. All reported data of structural and physical properties show
very similar features.

The low-temperature Mossbauer data for the perovskite Sr$_{2}$LaFe$_{3}$O$%
_{8.94}$ studied in \cite{battle} are consistent with a $2:1$ ratio of the
two types of $Fe$ cation with electronic characteristics fairly close to $%
Fe^{3+}$ and $Fe^{5+}$, respectively. The Neutron-Powder-Diffraction data at 
$5K$, show clear evidence that the low temperature phase exhibits an
antiferromagnetic ordering\cite{battle}. More precisely, there is
antiferromagnetic ordering among subcells of ferromagnetically-ordered
spins. At room temperature only is observed an average value of the charge,
which demonstrates the existence of a fast electron transfer.

So, upon heating the compound passes from a mixed-valence phase through a
first-order transition{\bf \ }to the paramagnetic average-valence phase.
Experimental data \cite{JCGFZ} for the specific heat show two peaks, the
highest of which seems to be related to a first-order phase transition. The
smaller peak, on the other hand, disappears as one fills in more oxygen
vacancies in the sample, as was shown by the data of ref.\cite{WLM} for the
sample Sr$_{2}$LaFe$_{3}$O$_{9}$, where the specific heat only shows one
(large) peak at about the same temperature as in \cite{JCGFZ}.

The inverse magnetic susceptibility for Sr$_{2}$LaFe$_{3}$O$_{8.94}$ shows%
\cite{battle} an abrupt change at about $200K$ with a minimum indicative of
a transition to antiferromagnetic ordering; and above this temperature the
susceptibility is almost field-independent; while below there is a
pronounced dependence on the magnetic field which appears to be an intrinsic
property of the phase, and suggests the possibility of a (weak){\bf \ }%
ferromagnetism, probably due to spin mis-alignment. The same sharp jump in
the magnetic susceptibility is reported in refs.\cite{JCGFZ} and \cite{WLM}.

According to previous structural studies\cite{battle}, this compound
exhibits an almost cubic perovskite structure, the rhombohedral distortion
being very small and not detectable by X-Ray Diffraction analysis. The
magnetic structure was determined from the Neutron-Diffraction data\cite
{battle} and is shown in figure 1.

The coupling between iron cations is of a superexchange-type via oxygen
anions. However, the usual approximation consists in treating the system as
that of iron cations interacting via an effective exchange integral. More
precisely, in our case the coupling between two $Fe^{3+}$ ions is
antiferromagnetic with the exchange constant denoted henceforth by\footnote{%
The subscripts $a$ and $f$ refer to antiferromagnetic and ferromagntic
orderings, respectively.} $J_{a}$, while that between $Fe^{3+}$ and $Fe^{5+}$
is ferromagnetic with the constant $J_{f}$, while there is no coupling
between two cations $Fe^{5+}$, as this would occur along the diagonal of the
(slightly deformed) elementary cube, and thus would be negligible in
comparison with the other two couplings (see figure 1). In our particular
case, these exchange integrals are obtained by fitting the experimental data
with a model of localized spins with two kinds of nearest-neighbor
interactions.

Accordingly, we consider a model that describes a system of $6$ localized
spins, four of $\frac{5}{2}$ and two of $\frac{3}{2}$, based on a
Dirac-Heisenberg Hamiltonian with antiferromagnetic and ferromagnetic
couplings $J_{a},J_{f}$, respectively and taking into account the
corresponding biquadratic interactions, denoted by $j_{a},j_{f}$. The latter
are known\cite{Bean} to induce a first-order transition because of the
strong change of the exchange interactions as a function of the interatomic
distances, and also the associated change in the elastic energy of the
material. The characteristic and very significant feature of such
magnetic-transformation mechanism is the sharp change of the elastic
constants of the crystal in the absence of any volume changes in the region
of the antiferromagnetic transformation\cite{Bean}. In our case, the
exchange-induced distortion of the crystal is of an order of magnitude too
small\cite{battle},\cite{JCGFZ} to be responsible for the change of the
order of the magnetic transition. In fact, it was pointed out in \cite
{Anderson1} and then shown in \cite{Huang} that the usual superexchange
mechanism is fully capable of explaining the origin and magnitude of the
biquadratic exchange interaction. Also Nagaev (\cite{Nagaev}, section 2.7)
studied the interplay between Dirac-Heisenberg and different kinds of
non-Dirac-Heisenberg interactions and the role of the latter in changing the
order of the magnetic transition viz in MgO, MnO and NiO. Moreover, since
the spins in our case are much larger than $\frac{1}{2}$, more excited spin
states start to participate in increasing the entropy of the system, and the
two-level-system approximation no longer holds as the temperature increases.
Therefore, in order to account for this effect, we must add higher-order
terms to the Dirac--Heisenberg Hamiltonian. Accordingly, we shall add
nearest-neighbor biquadratic interactions which, from the physical point of
view could account for the effect of charge disproportionation occurring in
our compound. Next, we show that indeed the biquadratic contributions to the
exchange energy account for the first-order magnetic transition occurring in
Sr$_{2}$LaFe$_{3}$O$_{9}$, and which is reflected by a sharp jump in the
magnetic susceptibility and large peak in the specific heat. Consistently,
we also predict an anomalous behavior for the sublattice magnetizations,
i.e. the jump down to zero in the vicinity of the N\'{e}el temperature.
Indeed, we obtain a good fit of the experimental data on the specific heat
and the magnetic susceptibility in both the ordered and disordered phases
for the exchange interactions $\frac{J_{a}}{k_{B}}\simeq 26K,\;\frac{J_{f}}{%
k_{B}}\simeq 6.5K$ ; and the corresponding biquadratic exchange integrals $%
j_{a}\simeq 0.11\times J_{a}$ and $j_{f}\simeq 0.08\times J_{f}$.

Note that the antiferromagnetic exchange integral $J_{a}$ between two iron
ions $Fe^{3+}$ in perovskite compounds such as LaFeO$_{3}$ was also
estimated by Anderson \cite{Anderson2} who found $J_{a}\simeq 26K$, and a
close value, $24K$, was also found by Grenier et al.\cite{Grenier et al.} in
CaFe$_{2}$O$_{5}$ with similar perovskite structure.

Next, we compute the magnon spectrum within the approach of linear spin-wave
theory. The spectrum contains a gapless antiferromagnetic mode and two
gapped ferromagnetic branches. Unfortunately, to the best of our knowledge,
there are so far no experimental data on the Inelastic-Neutron Scattering
for the present powder sample. However, to account for the anomalous
behavior of the sublattice magnetizations near the transition point, it
turns out that non-linear spin-wave corrections are necessary.

\section{Theory}

\paragraph{Hamiltonian}

The unit cell of our system is sketched in figure 1, and the
Dirac-Heisenberg Hamiltonian for such system can be written as 
\begin{equation}
H_{DH}=-J_{f}%
\mathop{\displaystyle \sum }%
\limits_{\left\langle i,j\right\rangle }%
\mathop{\displaystyle \sum }%
\limits_{\alpha ,\beta }{\bf s}_{i\alpha }\cdot {\bf S}_{j\beta }+J_{a}%
\mathop{\displaystyle \sum }%
\limits_{\left\langle i,j\right\rangle }%
\mathop{\displaystyle \sum }%
\limits_{\alpha ,\beta }{\bf S}_{i\alpha }\cdot {\bf S}_{j\beta }
\label{mf1}
\end{equation}
where henceforth ${\bf S},{\bf s}$ denote the spins $\frac{5}{2}$ of $%
Fe^{3+},$ and $\frac{3}{2}$ of $Fe^{5+}$, respectively. $%
\mathop{\displaystyle \sum }%
\limits_{\left\langle i,j\right\rangle }$ denotes the sum over all pairs of
nearest-neighbor sites $i,j$; each $Fe^{5+}$ ion having six $Fe^{3+}$ ions
as nearest-neighbors, while the nearest-neighbors of each $Fe^{3+}$ ion are
three $Fe^{3+}$ and three $Fe^{5+}$. The sum $%
\mathop{\displaystyle \sum }%
\limits_{\alpha ,\beta }$ runs over the six different atoms in the unit
cell, see figure 1.

Then to the Hamiltonian (\ref{mf1}) we add the following
non-Dirac-Heisenberg Hamiltonian of nearest-neighbor biquadratic
interactions 
\begin{equation}
H_{NDH}=-j_{f}%
\mathop{\displaystyle \sum }%
\limits_{\left\langle i,j\right\rangle }%
\mathop{\displaystyle \sum }%
\limits_{\alpha ,\beta }\left( {\bf s}_{i\alpha }\cdot {\bf S}_{j\beta
}\right) ^{2}-j_{a}%
\mathop{\displaystyle \sum }%
\limits_{\left\langle i,j\right\rangle }%
\mathop{\displaystyle \sum }%
\limits_{\alpha ,\beta }\left( {\bf S}_{i\alpha }\cdot {\bf S}_{j\beta
}\right) ^{2}  \label{mf2}
\end{equation}
Note that we have adopted the convention that all exchange couplings $%
J_{a},J_{f},j_{a},j_{f}$ are positive.

Within the mean-field (MF) approximation it turns out that in fact the
effect of adding these biquadratic interactions is to redefine the exchange
integrals $J_{a}$ and $J_{f}$, into temperature-dependent effective ones as
follows 
\begin{eqnarray}
J_{a} &\longrightarrow &J_{a}^{eff}=J_{a}+2j_{a}\left\langle S\right\rangle
^{2}  \label{Master equation} \\
J_{f} &\longrightarrow &J_{f}^{eff}=J_{f}+2j_{f}\left\langle S\right\rangle
\left\langle s\right\rangle  \nonumber
\end{eqnarray}
where $\left\langle S\right\rangle $ and $\left\langle s\right\rangle $ are
the spontaneous sublattice magnetizations on the sites $Fe^{3+}$ and $%
Fe^{5+} $, respectively. The same is also true within spin-wave theory, see
sect.4. However within the linear-spin-wave approximation the effective
couplings $J_{a}^{eff},J_{f}^{eff}$ are independent of temperature since in
this case the spins $S$ and $s$ are respectively substituted for the
sublattice magnetizations $\left\langle S\right\rangle $ and $\left\langle
s\right\rangle $.

As was demonstrated by Nagaev \cite{Nagaev}, all non-Dirac-Heisenberg
interactions, biquadratic in our case, change the order of the magnetic
transition from second to first. It is readily seen in (\ref{Master equation}%
) that the effective exchange integrals decrease in magnitude with
decreasing magnetizations because of a decrease in the order parameters $%
\left\langle S\right\rangle $ and $\left\langle s\right\rangle $. Inversely,
a decrease in the effective exchange integrals in turn leads to a decrease
in the magnetizations, i.e. there is a positive feedback effect. This brings
about the change in the order of the transition as the ratio, $\frac{j}{J}$
in the present case, between the non-Dirac-Heisenberg and the
Dirac-Heisenberg interaction reaches a certain critical value.

Henceforth, to avoid writing cumbersome formulae, we shall only give the
expressions for physical quantities derived from the Hamiltonian (\ref{mf1}%
), but keeping in mind that all calculations are done using the effective
exchange integrals $J_{a}^{eff},J_{f}^{eff}$ defined in (\ref{Master
equation}).

\paragraph{Spontaneous magnetizations}

Within the MF approximation the order parameters $\left\langle
S\right\rangle $, $\left\langle s\right\rangle $ are found to satisfy the
following coupled self-consistent equations

\begin{equation}
\begin{array}{l}
\left\langle S\right\rangle =B_{S}\left( z\beta \left[ J_{a}\left\langle
S\right\rangle +J_{f}\left\langle s\right\rangle \right] \right) \\ 
\\ 
\left\langle s\right\rangle =B_{s}\left( 2z\beta J_{f}\;\left\langle
S\right\rangle \right)
\end{array}
\label{mf4}
\end{equation}
where $\beta =\frac{1}{k_{B}T}$, and $B_{j}(x)$ is the usual Brillouin
function 
\[
B_{j}(x)=(j+\frac{1}{2})\coth (j+\frac{1}{2})x-\frac{1}{2}\coth \frac{x}{2} 
\]
and $z=3$, i.e. half the number of nearest neighbors of an atom.

Upon making the substitutions (\ref{Master equation}) the self-consistent
equations (\ref{mf4}) become more complicated and then they can be solved
only numerically. The corresponding solution is shown in solid lines in
figure 2, for the values of the exchange integrals obtained by fitting the
experimental data on the specific heat and magnetic susceptibility, see
sect.3 below.

\paragraph{Susceptibility}

We have computed the magnetic susceptibility both in the ordered and
paramagnetic phases following the generalized mean-field approach of Smart%
\cite{Smart}. For this purpose, we use the fact that the spontaneous
magnetizations given by eqs.(\ref{mf4}) exhibit the following
antiferromagnetic-ferromagnetic arrangement at $T<T_{N}$ (see figure 1)

\begin{equation}
\left\langle {\bf S}_{2}\right\rangle =-\left\langle {\bf S}%
_{3}\right\rangle =-\left\langle {\bf S}_{5}\right\rangle =\left\langle {\bf %
S}_{6}\right\rangle ,\quad \left\langle {\bf s}_{1}\right\rangle
=-\left\langle {\bf s}_{4}\right\rangle  \label{mf6}
\end{equation}
Thereby we obtain the following expressions for the transverse magnetic
susceptibility,

\begin{equation}
\chi _{\perp }{\bf =}\alpha \times \frac{J_{f}\left( 2\left\langle
S\right\rangle +\left\langle s\right\rangle \right) ^{2}+2J_{a}\left\langle
S\right\rangle \left\langle s\right\rangle }{2zJ_{f}J_{a}\;\left\langle
S\right\rangle ^{2}}  \label{mf7}
\end{equation}
and parallel susceptibility\footnote{%
Here the prime stands for the derivative of the Brillouin function $B(x)$%
with respect to $x$.}{\bf \ } 
\begin{equation}
{\bf \chi }_{\mid \mid }=\alpha \times \frac{2T\cdot \left( 2B_{S}^{^{\prime
}}(y_{0}^{S})+B_{s}^{^{\prime }}(y_{0}^{s})\right) +2z\cdot \left(
J_{a}+4J_{f}\right) B_{s}^{^{\prime }}(y_{0}^{s})B_{S}^{^{\prime
}}(y_{0}^{S})}{T^{2}+zJ_{a}T\cdot B_{S}^{^{\prime
}}(y_{0}^{S})-2z^{2}J_{f}^{2}\ \cdot B_{s}^{^{\prime
}}(y_{0}^{s})B_{S}^{^{\prime }}(y_{0}^{S})}  \label{mf8}
\end{equation}
where 
\[
y_{0}^{s}=\left( 2zJ_{f}\left\langle S\right\rangle \right) \beta ,\qquad \
y_{0}^{S}=z\left( J_{f}\left\langle s\right\rangle +J_{a}\left\langle
S\right\rangle \right) \beta 
\]
and have introduced the conversion coefficient $\alpha =\mu _{0}(g\mu
_{B})^{2}N_{c}/k_{B}$. With $g=2$, and $N_{c}=\frac{N_{A}}{2},N_{A}$ being
the Avogadro number, as there are two molecules of the sample in the unit
cell.

It is worthwhile to note that the transverse susceptibility (\ref{mf7}) is a
decreasing function of temperature.

Now, as we are dealing with powder sample, the total magnetic susceptibility
below the N\'{e}el temperature is given by 
\[
\chi (T)=\frac{1}{3}{\bf \chi }_{\mid \mid }+\frac{2}{3}\chi _{\perp } 
\]

Next, in the paramagnetic phase the total magnetic susceptibility is found
to be 
\begin{equation}
{\bf \chi }_{PM}=\alpha \times \frac{2}{z}\cdot \frac{y+2x+J_{a}-4J_{f}}{%
xy+J_{a}x-2J_{f}^{2}}\,  \label{mf9}
\end{equation}
where 
\[
x=\frac{3k_{B}T}{z}\frac{1}{s(s+1)},\qquad y=\frac{3k_{B}T}{z}\frac{1}{S(S+1)%
} 
\]
One can check that as $T\rightarrow T_{N}$, $B_{j}^{^{\prime
}}(x)\rightarrow \frac{j(j+1)}{3}$, so that eq.(\ref{mf8}) reduces to eq.(%
\ref{mf9}), and at $T=T_{N}$, we get $\chi _{\perp }={\bf \chi }_{\mid \mid
}={\bf \chi }_{PM}$.

\paragraph{Entropy}

Within the foregoing approach the entropy (per atom) of the system reads 
\begin{equation}
\frac{S(T)}{N_{c}}=2\left[ \log \left( C_{\sigma }(\beta A)\right) -\beta
A\left\langle s\right\rangle \right] +4\left[ \log \left( C_{S}(\beta
B)\right) -\beta B\left\langle S\right\rangle \right]  \label{entropy}
\end{equation}
where $A=2zJ_{f}\left\langle S\right\rangle $, $B=z\left[ J_{a}\left\langle
S\right\rangle +J_{f}\left\langle \sigma \right\rangle \right] $, and $%
C_{j}(x)=\frac{\sinh \left[ (2j+1)\frac{x}{2}\right] }{\sinh (\frac{x}{2})}$%
, whose derivative is the Brillouin function given earlier.

When plotted as a function of temperature, for the exchange couplings
obtained below, the entropy (\ref{entropy}) increases with increasing
temperature up to the N\'{e}el point where it exhibits an abrupt jump
characteristic of a first-order transition, and then it saturates to a
constant given by the configurational entropy in the paramagnetic phase.

\paragraph{Specific heat}

The specific heat within the same approximation is given by\footnote{$%
\partial _{T}$ stands for the derivative with respect to temperature.} 
\begin{equation}
\frac{C_{v}}{N_{c}}=-2z\cdot \left( J_{a}\cdot \left\langle S\right\rangle
\partial _{T}\left\langle S\right\rangle +J_{f}\cdot \left[ \left\langle
s\right\rangle \partial _{T}\left\langle S\right\rangle +\left\langle
S\right\rangle \partial _{T}\left\langle s\right\rangle \right] \right)
\label{Cv}
\end{equation}
for $T\lesssim T_{N}$.

Above $T_{N}$ the mean-field approximation yields a zero specific heat since
in this temperature range the order parameters and thereby the free energy
vanish, see figure 4.

\section{Results and Discussion}

The Dirac-Heisenberg model including the biquadratic interactions studied
here has been written for the ideal compound Sr$_{2}$LaFe$_{3}$O$_{9}$ with
no oxygen vacancies. On the other hand, the compound with the closest
composition to the latter was studied experimentally by Wang et al. in\cite
{WLM}. Therefore, we believe that our theory is more suitable for fitting
the experimental data of ref.\cite{WLM} than those of \cite{battle} or \cite
{JCGFZ}, especially from the quantitative point of view. However, we also
obtain good qualitative agreement with the authors of refs.\cite{battle}, 
\cite{JCGFZ}.

A reasonable (numerical) fit to the experimental data\cite{WLM} on the
magnetic susceptibility and specific heat given in figures 3 and 4, led to
the following values of the exchange integrals and biquadratic interactions 
\begin{eqnarray}
\frac{J_{a}}{k_{B}} &\simeq &26K,\qquad \frac{J_{f}}{k_{B}}\simeq 6.5K
\label{couplings} \\
j_{a} &\simeq &0.11\times J_{a},\quad j_{f}\simeq 0.08\times J_{f}  \nonumber
\end{eqnarray}

As was mentioned in the introduction, the value of the antiferromagnetic
exchange integral $J_{a}$ found here is in agreement with the ones obtained
by Anderson \cite{Anderson2}, i.e. $J_{a}\simeq 26K$ in LaFeO$_{3}$, or by
Grenier et al.\cite{Grenier et al.}, that is $24K$ in CaFe$_{2}$O$_{5}$.

For these couplings we find that the sublattice magnetizations exhibit a
sharp drop at the N\'{e}el temperature indicating that the transition is of
first order, see figure 2. This behavior is also reflected in the magnetic
susceptibility, which exhibits a sharp jump at the transition, in agreement
with the susceptibility measured by Wang et al.\cite{WLM} and also with the
one reported by Zhou et al.\cite{JCGFZ}, see figure 3. We see that the
susceptibility curve obtained within the mean-field theory fits both
qualitatively and quantitatively to the one measured by Wang et al.\cite{WLM}%
, however the authors of ref.\cite{JCGFZ} obtain a larger jump at the
critical temperature.

In addition, the jump in the entropy (\ref{entropy}) is indicative of a
first-order transition, and yields the latent heat released by the system at
the transition. Accordingly, for the exchange integrals found above, we find
that the contribution of magnetic excitations to the latent heat is given by
the product of the jump in the corresponding entropy at the critical
temperature and the latter, i.e. $Q(T_{N})=T_{N}\times \Delta S\simeq
2.9\;KJ/mole$. The experimental enthalpy found by Zhou et al.\cite{JCGFZ} is 
$\Delta H\simeq 3.7\;KJ/mole$, which is the total enthalpy of the system.

Consequently, the specific heat obtained within the mean-field approximation
(\ref{Cv}) diverges at the N\'{e}el temperature. In the magnetically ordered
phase and around the transition we obtain a good fit of this to the
experimental curve obtained by Wang et al.\cite{WLM}, see figure 4. As is
well known the mean-field theory cannot be correct above the N\'{e}el
temperature, for it predicts the absence of short-range order. In
particular, in our case the specific heat computed within mean-field
approximation drops to zero for $T>T_{N}$, since then the order parameters
vanish, i.e. $\left\langle S\right\rangle =$ $\left\langle s\right\rangle =0$%
. Alternatively we have computed the contribution to the specific heat (\ref
{Cv}) in the paramagnetic phase using the appraoch of high-temperature
expansion, but this does not yield a significant contribution. We have also
taken into account that we are only dealing with the contribution of
magnetic excitations to the transition while leaving out the lattice
component.

To derive the Curie-Weiss law for the magnetic susceptibility in the
paramagnetic phase, we note that the denominator in (\ref{mf9}) is quadratic
in temperature, and thus leads to a hyperbolic function of temperature, as
in the case of a ferrimagnet. However the high temperature asymptote to the
hyperbola does have the Curie-Weiss law form, that is 
\begin{equation}
\chi =\frac{C_{a}}{T-\theta _{a}}  \nonumber
\end{equation}
with 
\[
C_{a}\simeq 12,\quad \theta _{a}\simeq -228{}K 
\]
in fair agreement with the experimental result\cite{battle}, \cite{JCGFZ} 
\[
C=11.4,\qquad \theta =-250{}K 
\]

In the paramagnetic phase, we can also estimate the average magnetic moment
corresponding to the average valence of iron. Indeed, the relationship
between the magnetic moment per atom and susceptibility 
\[
\mu =\sqrt{8\times \chi (T-\theta _{a})}=2.83\times \sqrt{\frac{C_{a}}{3}} 
\]
yields for iron 
\[
\mu (Fe)=2.83\times \sqrt{\frac{C_{a}(Fe)}{3}}\simeq 5.66\mu _{B} 
\]
which agrees with the value found by Battle et al.\cite{battle}, that is $%
5.55\mu _{B}$.

\section{spin wave theory}

\paragraph{Spectrum}

Within the linear spin-wave theory based on the Holstein-Primakoff
representation\cite{holstein/primakoff} of spin operators ($\frac{1}{S}$
expansion), we find three doubly-degenerate magnon branches, as shown in
figure 5. The explicit expressions of the corresponding magnon energies are
rather messy and we omit writing them here. Nonetheless, around the point $%
\Gamma ,k=(0,0,0),$ located at the center of the Brillouin zone we can
write, 
\begin{eqnarray}
\hbar \omega _{1}(k) &=&\sqrt{\rho _{xy}^{1}\cdot (k_{x}^{2}+k_{y}^{2})+\rho
_{z}^{1}\cdot k_{z}^{2}}  \label{spectrum} \\
\hbar \omega _{2}(k) &=&\sqrt{\Delta _{1}+\rho _{xy}^{2}\cdot
(k_{x}^{2}+k_{y}^{2})+\rho _{z}^{2}\cdot k_{z}^{2}}  \nonumber \\
\hbar \omega _{3}(k) &=&\sqrt{\Delta _{2}+\rho _{xy}^{3}\cdot
(k_{x}^{2}+k_{y}^{2})-\rho _{z}^{3}\cdot k_{z}^{2}}  \nonumber
\end{eqnarray}
where the gaps $\Delta _{1}$ and $\Delta _{2}$ are given by 
\begin{eqnarray*}
\Delta _{1} &=&\frac{9J_{f}}{4}\left[ 2J_{a}\;sS+J_{f}(2S+s)^{2}\right] \\
\Delta _{2} &=&\frac{9J_{f}}{4}\left( 2J_{a}sS+J_{f}s^{2}\right)
\end{eqnarray*}
and the spin stiffness coefficients $\rho _{xy}^{1},\rho _{z}^{1}$, etc.,
are (cumbersome) functions of the exchange integrals. Recall that the
exchange integrals $J_{a}$ and $J_{f}$ must be redefined using eq.(\ref
{Master equation}) and taking into account the fact that in the
linear-spin-wave approximation the sublattice magnetizations $\left\langle
S\right\rangle $ and $\left\langle s\right\rangle $ are replaced by their
nominative values $S=\frac{5}{2}$ and $s=\frac{3}{2}$, respectively.

It is seen in figure 5 that the first of the spectrum branches, the lowest
curve, is gapless and of antiferromagnetic type. Whereas, the second and
third branches, the upper ones, represent gapped ferromagnetic modes. We
have assumed here that the anisotropy is too small to produce a gap in the
magnon spetrum at temperatures the latter is obtained.

Therefore, the spectrum above shows that we have antiferromagnetic ordering
at low temperature, together with a ''weak ferromagnetic ordering'' that
starts propagating upon heating. As discussed in the introduction, it should
be interesting to compare our results for the magnon spectrum with the
experimental data, were Inelastic-Neutron Scattering measurements possible
on the compound Sr$_{2}$LaFe$_{3}$O$_{9}$. This would also allow us to
compare the spin stiffiness coefficients we have found here with the
experimental ones, and then determine the effective magnetic moments of $%
Fe^{3+}$ and $Fe^{5+}$ at zero temperature, and thereby estimate their
reduction by quantum fluctuations.

\paragraph{Brillouin zone}

The Brillouin zone is a $3D-$hexagon (see figure 5) whose edges are defined
by 
\[
\frac{-\pi }{\sqrt{6}}\leq k_{z}\leq \frac{\pi }{\sqrt{6}},\;\frac{-2\pi }{%
\sqrt{3}}\leq k_{y}\leq \frac{2\pi }{\sqrt{3}},\;-\left| \frac{4\pi }{3}-%
\frac{\left| k_{y}\right| }{\sqrt{3}}\right| \leq k_{x}\leq \left| \frac{%
4\pi }{3}-\frac{\left| k_{y}\right| }{\sqrt{3}}\right| . 
\]

\paragraph{Thermodynamic quantities}

The ground-state energy per site is given by 
\begin{equation}
\frac{E_{g}}{N_{c}}=H_{0}+\frac{1}{N_{c}}\frac{\sqrt{18}}{2}\sum_{\nu =1}^{3}%
\displaystyle \int %
\limits_{B.Z.}\frac{d^{3}k}{(2\pi )^{3}}\;\hbar \omega _{\nu }(k)  \nonumber
\end{equation}
where 
\[
H_{0}=-zJ_{a}\cdot S(S+1)-zJ_{f}\cdot \left[ S(s+1)+s(S+1)\right] \simeq
-0.16\;\,eV 
\]
and $\frac{E_{g}}{N_{c}}\simeq -0.12\;\,eV$.

Next, within linear spin-wave theory we have studied the temperature
dependence of the internal energy, specific heat, and sublattice
magnetizations. We have found that at low temperatures, the internal energy
behaves as $T^{4}$, and that the specific heat behaves as $T^{3}$, which is
consistent with an antiferromagnetic ordering at low temperatures.

On the other hand, as discussed in the introduction, the linear spin-wave
theory turns out to be a poor approximation in the vicinity of the
transition, as long as sublattice magnetizations are concerned. Indeed we
find that $\left\langle S\right\rangle $ and $\left\langle s\right\rangle $
decrease linearly with temperature as this approaches the N\'{e}el point,
and vanish at different temperatures, see figure 2. There we also see that
the linear-spin-wave theory yields, at zero temperature, quantum corrections
to the magnetic moments of $Fe^{3+}$ ions but no corrections to those of $%
Fe^{5+}$ ions. Therefore, we think that further non-linear corrections
should be taken into account so as to obtain the correct temperature
dependence of the sublattice magnetizations especially as the temperature
approaches the N\'{e}el point.

\section{Conclusion}

We have modelled the first-order magnetic transition occurring in the
perovskite $Sr_{2}LaFe_{3}O_{9}$ using a Dirac-Heisenberg Hamiltonian
including (nearest-neighbor) biquadratic interactions, whose origin could be
related with the disproportionation of iron. We have been able, by fitting
the experimental magnetic susceptibility and specific heat, to estimate the
superexchange integrals for the antiferromagnetic and ferromagnetic
interactions, as well as the biquadratic ones. The first of these integrals
is consistent with the results of previous work on kindred compounds.

Elastic-Neutron Scattering will be performed on this compound for checking
the anomalous behavior of the sublattice magnetizations predicted here.
Unfortunately, it is difficult to check up on the magnon spectrum we have
computed. Nevertheless, the latter does confirm the magnetic structure
determined from Neutron-Diffraction data by Battle et al.\cite{battle} in
the compound $Sr_{2}LaFe_{3}O_{8.94}$.

Non-linear spin-wave calculations and Monte Carlo simulations are still
under investigation. Finally, it should be very instructive to understand
more in detail the effect of iron disproportionation, or more generally the
charge redistribution, on the magnetic ordering.

\section{Acknowledgments}

P.K. would like to acknowledge the kind hospitality extended to him by the
Laboratoire de Physique Th\'{e}orique during his stay. We would like to
thank M.L. Kulic, A.I.Buzdin, J.P.Brison and D.Foerster for helpful
discussions; and J.Leandri for checking our fit of the exchange constants
using the MINUIT algorithm.

{\Large Figure captions}

\begin{itemize}
\item  Figure 1: The crystal and magnetic structure of Sr$_{2}$LaFe$_{3}$O$%
_{9}\cite{battle}$. The large circles represent $Fe^{3+}$ and the medium
ones represent $Fe^{5+}$, with the arrows indicating the size and
orientation of their spins; the smallest circles stand for intermediate
oxygen atoms. The $2Sr^{2+}:La^{3+}$ cations which are disorderly located in
the cubic centers have been omitted for clarity. The hexagonal unit cells of
the two structures are commensurate, but the unit cell of the crystal
structure is a triple one, while that of the magnetic structure is a
primitive one, for the rhombohedral translations are absent in the magnetic
structure. Each unit cell contains two formulae of $Sr_{2}LaFe_{3}O_{9}$,
hence six iron ions, which form six magnetic sublattices below the ordering
temperature.
\end{itemize}

\vspace*{1cm}

\begin{itemize}
\item  Figure 2: The solid curves represent the temperature dependence of
the sublattice magnetizations predicted by mean-field theory including
biquadratic interactions. The upper (solid) curve represents the
magnetization $\left\langle S\right\rangle $ of the $Fe^{3+}$ ions of spin $%
\frac{5}{2}$, and the lower curve is the magnetization $\left\langle
s\right\rangle $ of $Fe^{5+}$ of spin $\frac{3}{2}$, obtained for the
exchange integrals given in eq.(\ref{couplings}) in the text.

The dashed curves represent the corresponding sublattice magnetizations
predicted by the linear-spin-wave theory, up to a temperature of $120K$.
\end{itemize}

\vspace*{1cm}

\begin{itemize}
\item  Figure 3: The curve in balls represents the experimental magnetic
susceptibility of ref.\cite{WLM}, and the one in triangles is the
suscpetibility measured by Zhou et al.\cite{JCGFZ}. The solid line is the
magnetic susceptibility obtained from MF theory including biquadratic
interactions.
\end{itemize}

\vspace*{1cm}

\begin{itemize}
\item  Figure 4: The curve in balls is the experimental specific heat of ref.%
\cite{WLM}, and the solid line is our theoretical result of MF theory
including biquadratic interactions.
\end{itemize}

\vspace*{1cm}

\begin{itemize}
\item  Figure 5: Plot of the magnon spectrum along the path $Z\Gamma
XM\Gamma Y$ indicated in the Brillouin zone shown as inset.
\end{itemize}

\end{document}